\documentclass{acm_proc_article-sp}
\usepackage{multirow}
\usepackage{listings}
\usepackage{seqsplit}
\usepackage{xcolor}
\usepackage{soul}       
\usepackage{calc}                 
\usepackage[romanian]{babel}
\usepackage{combelow}
\begin{document}

\title{Summoning Demons}
\subtitle{The Pursuit of Exploitable Bugs in Machine Learning}

\numberofauthors{6} 
\author{
\alignauthor Rock Stevens
\alignauthor Octavian Suciu
\alignauthor Andrew Ruef
\and
\alignauthor Sanghyun Hong
\alignauthor Michael Hicks
\alignauthor Tudor Dumitra\cb{s}
\and
\affaddr{University of Maryland, College Park}
}

\newcommand{\TODO}[2][]{\noindent\colorbox{lime}{
    \color{red}
    \parbox{\minof{.97\hsize}{\widthof{#1} * \real{1.15} + \widthof{ #2}}}{\textbf{#1} #2}
    \par}
}

\sethlcolor{lime}
\newcommand{\todo}[2][]{
    \textcolor{red}{\hl{\textbf{#1} #2}}
}

\sloppy

\maketitle

\begin{abstract}
\noindent 
Governments and businesses increasingly rely on data analytics and machine learning (ML) for improving their competitive edge in areas such as consumer satisfaction, threat intelligence, decision making, and product efficiency. 
However, by cleverly corrupting a subset of data used as input to a target's ML algorithms, an adversary can perturb outcomes and compromise the effectiveness of ML technology.
While prior work in the field of adversarial
machine learning has studied the impact of input
manipulation on correct ML algorithms, we consider the
exploitation of bugs in ML implementations. In this paper,
we characterize the attack surface of ML programs,
and we show that malicious inputs exploiting implementation
bugs enable strictly more powerful attacks than the
classic adversarial machine learning techniques. We propose
a semi-automated technique, called steered fuzzing,
for exploring this attack surface and for discovering exploitable
bugs in machine learning programs, in order to
demonstrate the magnitude of this threat. 
As a result
of our work, we responsibly disclosed five vulnerabilities,
established three new CVE-IDs, and illuminated a
common insecure practice across many machine learning
systems.
Finally, we outline several research directions for
further understanding and mitigating this threat.


\end{abstract}


\section{Introduction}
\label{sec:intro}
Governments and businesses increasingly employ data analytics to improve their competitive edge. For example, the United States Environmental Protection Agency has outlined its vision for leveraging machine learning (ML) to improve their everyday operations \cite{epa.gov}. IBM offers businesses a platform for conducting sentiment analysis to gauge their effectiveness within a target audience \cite{ibm_sentiment}. OpenDNS uses ML to automate protection against known and emerging threats \cite{opendns}.
Machine learning allows these organizations to extrapolate trends from massive data sets, of often uncertain provenance.

However, ingesting unfiltered, public information into data analytic engines also introduces a threat, as miscreants can corrupt eventual inputs to ML algorithms to bias their outputs.
Cretu et al.~\cite{cretu2008casting} discussed the importance of ``casting out demons,'' or sanitizing
the training datasets for safe machine learning ingestion.
Research on \emph{adversarial machine learning} \cite{lowd2005adversarial, miller2014adversarial, barreno2010security, biggio2013data,goodfellow2014explaining}
has explored various attacks against ML \emph{algorithms}, with a focus on skewing their outputs through malicious perturbations to the input data.

In this paper, we discuss another attack vector: ML algorithm \emph{implementations.} Like all software, ML algorithm implementations have bugs and some of these bugs could affect learning tasks. Thus, attacks can construct malicious inputs to ML algorithm implementations that exploit these bugs. Indeed, such attacks can be more powerful than traditional adversarial machine learning techniques.
For example, a memory corruption
vulnerability could allow an adversary to corrupt
the entire feature matrix, not just the entries that correspond
to adversary-controlled inputs. More generally,
bugs in the cost function, minimization algorithm, model
representation, prediction or clustering steps, could allow
an adversary to arbitrarily skew learning outcomes or
to initiate a denial of service attack.

While considerable efforts have been devoted to discovering software vulnerabilities and mitigating the impact of exploits, these generally focus on bugs that allow the adversary to subvert the targeted system, e.g. by executing arbitrary code or by achieving privilege escalation.
In contrast, adversaries attacking an ML system are interested in bugs that allow them to induce mispredictions, misclustering, or to suppress outputs.
Such logic bugs are difficult to discover using existing tools.

As a first step toward understanding and mitigating this threat, we characterize the attack surface of ML programs, which derives from a general architecture that many ML algorithms share, and we identify decision points whose outcome we may corrupt.
We discuss how bugs around those decision points could be exploited and the potential outcomes of these exploits.
We also propose a semi-automated technique called \emph{steered fuzzing} for finding and exploiting ML implementation bugs.
We wrap important decision points from the ML architecture with instrumented code to convert a logical failure of the algorithm (e.g. misprediction) into a crash that can be detected by a fuzzing tool~\cite{DBLP:journals/cacm/MillerFS90}, which generates test cases and records program exceptions on these inputs.
We then apply a
coverage-based fuzzing tool, American Fuzzy Lop~\cite{afl}, to \emph{summon demons}, i.e. to automatically discover 
inputs that mislead the ML algorithms by exploiting bugs in their implementation.

We utilize this technique to discover attacks against OpenCV~\cite{opencv} and Malheur~\cite{malheur}, two open source ML implementations. As an example, we started fuzzing with a seed image that was recognized as having a face by OpenCV. We added a logic branch that crashed on non-recognition. Steered fuzzing then proceeded to generate a mutant input that was clearly still a face, and yet was not recognized. This exploit relies on a bug in the rendering library used by OpenCV, which allows for incorrect rendering of input images.
In total, we found seven bugs: three in OpenCV, two in Malheur, one in Scikit-Learn, and one in {\tt libarchive} (used by Malheur). Of these,
three were assigned a CVE-ID; only one was not exploitable.

In summary, this paper makes three contributions. First, we explore the attack surface of ML implementations, as compared to ML algorithms, highlighting potential attack vectors and impact on various components within these systems. Second, we introduce a novel technique for exploiting ML bugs to corrupt classification outcomes and the data provided to ML systems from benign sources. This technique is possible through \emph{steered fuzzing}, which expands upon existing fuzzing techniques for discovering bugs in  software applications.
Finally, we discover several new ML implementation bugs in important open-source software; our work has led to these bugs being patched.


\section{Problem Statement}

We consider an exploit to be a piece of code aiming to subvert the intended functionality of software.
Limiting our scope to machine learning, an exploit would be designed with the goal of corrupting the outputs of programs or to inhibit their operation.
Such exploitable bugs may be present either in the core implementation of the ML algorithm or in libraries used for feature extraction or model representation. 

In terms of impact, we distinguish between three possible outcomes of successful exploits.
First, an exploit that causes specific instances to be assigned an incorrect label achieves \emph{mispredictions}.
Specifically, an exploit targeting the training phase results in \emph{poisoning} of the classifier, while one during testing could allow \emph{evasion}. 
Similarly, an exploit targeting a clustering algorithm may cause inputs to be placed in different clusters, resulting in \emph{misclustering}.
Because machine learning systems are often utilized as black boxes, it may be difficult to detect that the system has been compromised by using one of these exploits, as they typically have no other side effects besides skewing the learned model and cause the ML system to fail silently. 
An exploit may also result in \emph{denial of service}, e.g. by stopping data ingestion prematurely or by crashing the application to prevent it from providing any output. 
While easier to detect, such exploits may render the system temporarily unusable. 
Finally, a successful exploit could enable \emph{code execution} capabilities for the attacker. 
This category of exploits is arguably the most powerful since it gives the adversary full control over the underlying machine. 

In this paper, we address the problem of discovering exploitable vulnerabilities in machine learning \textit{implementations}.
The goals of our work are: (i) to provide a general ML architectural description, discussing possible attack vectors and their impact on different system components; (ii) to develop a semi-automated technique for discovering ML vulnerabilities by exploring this attack surface; and (iii) to demonstrate the magnitude of this threat by discussing several real vulnerabilities we discovered in popular ML systems. 

\textbf{Non-goals:} We do not address limitations of machine learning algorithms (the area of study in adversarial machine learning).
Instead, we aim to unearth implementation bugs as an orthogonal attack vector against ML systems.
Additionally, we do not aim to develop a fully automated technique for identifying these bugs. 
Instead, by relying on steered code instrumentation and program output manipulation, we are able to bootstrap existing fuzzing tools in order to discover bugs.

\subsection{Threat Model}
\label{sec:problem}

We consider an adversary who aims to subvert the execution of machine learning algorithms by exploiting bugs in algorithm implementations.
We assume that the adversary has access to the program's source code.
We also assume that the adversary controls some of the program's inputs, but is unable to prevent benign users from providing additional inputs.
These assumptions are realistic in many settings; for example, the machine learning techniques proposed for malware classification or clustering~%
\cite{DBLP:conf/sp/SchultzEZS01,
	DBLP:conf/nsdi/PerdisciLF10,
	DBLP:conf/icassp/DahlSDY13,
	malheur}
operate on inputs that come from many sources, including possible adversaries.
Additionally, much ML software is open source, e.g., OpenCV~\cite{opencv} and Scikit-Learn~\cite{sklearn}.

In searching for exploitable bugs, the adversary does not pursue the usual goals of vulnerability exploitation, e.g., gaining control over remote hosts, achieving privilege escalation, escaping sandboxes, etc.
Instead, the adversary's goal is to corrupt the outputs of machine learning programs using silent failures.

From a spectrum-of-control perspective, arbitrary code execution exploits represent the strongest means for achieving the adversary's goal, as such an exploit permits an adversary to manipulate all aspects of the target system.
However, the adversary may achieve her goals with less powerful logical exploits, e.g., targeting memory corruption bugs that allow modifying data in memory but do not enable code execution or bugs that trigger loss of precision in floating point computations.
Denial of service attacks could also prove beneficial for the attacker
and may, for example, be conducted by inducing early termination of the ML processing.
In some settings, the weaker attacks may be more attractive as they could allow the adversary to remain stealthy and bypass defense mechanisms.


\begin{figure*}[t]
\centering
\includegraphics[width=\textwidth]{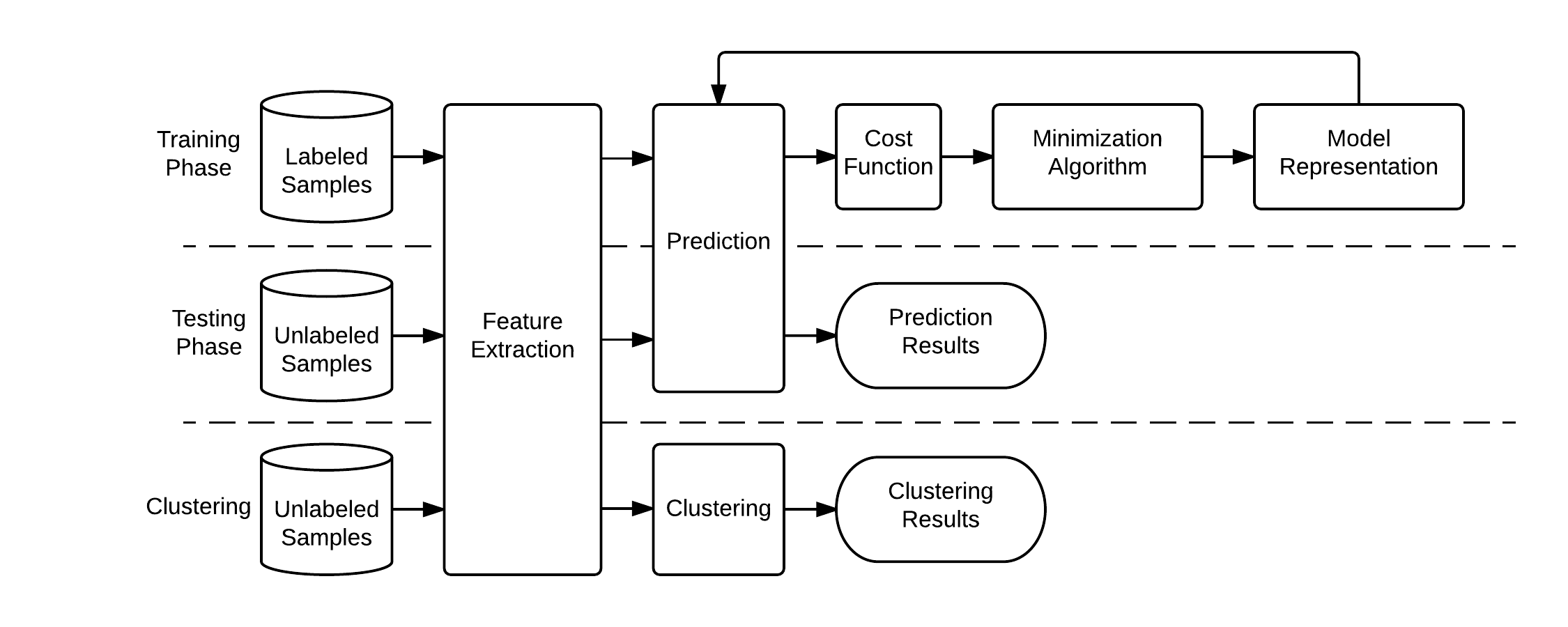}
\caption{General architecture of ML systems.}
\label{attacksurface}
\end{figure*}

\section{Attacking ML Implementations}
\label{sec:methods}
To begin exploring the vulnerabilities of machine learning applications, we must first understand their attack surface.
Enumerating the components of ML applications that an attacker may target allows us to reason about
where the bugs may be and what impact they may have.
We then build on this understanding and expand upon existing fuzzing techniques to create exploits for these bugs to induce
misclassifications (false negatives or false positives), incorrect clustering results, and denial of service.

\subsection{Machine Learning Architecture}
Machine learning algorithms vary in structure and design. The two main categories of learning algorithms are supervised and unsupervised. In supervised learning, the algorithm receives a set of labeled examples and computes a predictive model that fits the training examples. The predictive model is then used to classify new, unlabeled samples. In contrast, unsupervised learning relies solely on unlabeled examples with the goal of finding clusters of similar samples.
While there may not be a generic representation that fits all algorithms, some of the most popular supervised techniques are variations of \emph{iterative minimization algorithms}. For unsupervised learning, \emph{clustering} is one of the most prevalent classes of algorithms.
Figure \ref{attacksurface} presents the general flow of a learning algorithm, highlighting the key particularities of each phase. In this setting, the input samples are transformed into a feature matrix representation that serves as the input to the classifier. A common, but optional, practice is to normalize the features prior to feeding them to the algorithm. This involves feature scaling and standardization. In the training phase, the (normalized) features are applied onto the current model in order to obtain the perceived prediction.
The predictions are compared to the actual class labels using a cost function. The cost function output quantifies the distance between the current model and the ground truth. The model is then updated to reduce the cost through a minimization algorithm. This iterative process is repeated until the model becomes a sufficiently accurate representation of the ground truth. Upon convergence, the model is used to predict new class labels. In the testing phase, the unlabeled samples are transformed using the same feature extraction and normalization processes. The predicted class labels are obtained using the prediction function over the trained model. In clustering, the algorithm first performs feature extraction and normalization. Using a distance metric, the algorithm groups the samples into clusters that reflect the similarity between them.

\subsection{From Architecture to Attack Surface}
\label{sec:attack-surface}

\begin{table*}[t]
\centering
\begin{tabular}{|c|c|c|}
\hline
{\bf Component}             & {\bf Exploitation Techniques}                               & {\bf Impact}                        \\
\hline
\multirow{2}{*}{Feature Extraction}     & \multirow{2}{*}{Insufficient integrity checks} & Poisoning / Evasion / Misclustering, \\
& & Code execution, DoS   \\
\hline
Prediction             & Overflow / Underflow, NaN, Loss of Precision & Poisoning / Evasion \\
\hline
Cost Function          & Overflow / Underflow, NaN, Loss of Precision & Poisoning, DoS \\
\hline
Minimization Algorithm & Overflow / Underflow, NaN, Loss of Precision & Poisoning, DoS \\
\hline
Model Representation   & Loss of Precision                           & Poisoning / Evasion \\
\hline
Clustering             & Overflow / Underflow, NaN, Loss of Precision & Misclustering                 \\
\hline
\end{tabular}
\caption{Attack surface of ML algorithms.}
\label{tableattacksurface}
\end{table*}

We now discuss how attacks on each component in Figure \ref{attacksurface} may impact the overall functioning of the system. Table~\ref{tableattacksurface} summarizes the vectors and impact of attacks against the system components. A successful attack against one component may have ripple effects to others, either directly by transferring corrupted outputs to inputs, or indirectly via in-memory data structure corruption.

\vspace{-0.2in}
\paragraph{Feature extraction}
Feature extraction is the backbone for the integrity of the system. Every attack from an external source must exploit vulnerabilities in this component as it is the sole communication port between the internal components and the external environment. An attack targeting the feature extraction component results in a corruption of the information passed downstream.

Within the feature extraction component itself, an attack can target the input parsing algorithms and/or the integrity checks performed on the feature representation.
As shown in Section~\ref{sec:results}, such an exploit could result in mispredictions, misclustering, arbitrary code execution or DoS.
It is not always straightforward to define what is allowable input, or an allowable representation thereof. For example, in an image classification setting, a reasonable assumption would be to consider any renderable image as legitimate inputs. However, as detailed in Section \ref{fuzzing}, we found that most images that cause crashes in the OpenCV library are actually valid from a rendering perspective.

\vspace{-0.2in}
\paragraph{Prediction}
Attacks are also possible against the prediction component, directly influencing the labels predicted by the algorithm. This could occur in both the training and the testing stages. For example, the attack could exploit bugs related to floating point overflow, floating point underflow, or the use of not-a-number ({\tt NaN}) values. ML implementations typically compute logarithms and square roots. This makes them particularly susceptible to bugs caused by NaNs (potentially the result of an overflow or insufficient consistency checks). The effects of a NaN propagate throughout the remainder of the computation and can result in poisoning or classifier evasion.

\vspace{-0.2in}
\paragraph{Cost function and minimization algorithm}
The cost function computation and the minimization algorithm are iteratively applied in the training phase. A bug could result in incorrect cost estimates or model updates that cause the decision boundary to be shifted away from the optimal value. Additionally, a denial of service could be obtained if the model update does not trigger the termination condition in the iterative algorithm. If the cost function consistently results in a NaN for the training examples, the minimization algorithm stagnates indefinitely without updating the model.

\vspace{-0.2in}
\paragraph{Model representation}
The model representation could cause poisoning or evasion through loss of precision. Since the training and the testing phases of algorithms are typically performed separately, the model has to be stored and transferred from the one to the other. As discussed in Section~\ref{sec:results}, casting between \emph{float} and \emph{long} types can skew the model away, resulting in inaccurate predictions for the unlabeled samples.

\vspace{-0.2in}
\paragraph{Clustering}
In clustering, the algorithm itself or the distance metric can be manipulated using the same attack vectors as for supervised learning. This could result in a denial of service or misclustering. In complete misclustering, the clusters are completely misrepresented, while in selective misclustering the attack might result in a particular sample being placed in a different cluster.

\subsection{Discovery Methods} \label{discovery}
\emph{Fuzzing}~\cite{DBLP:journals/cacm/MillerFS90} is a popular method for bug discovery.
A fuzzing tool tests a program using randomly generated inputs, which are often invalid or unexpected by the implementation, and records program exceptions or failures.
In security,
fuzzing has been employed to identify crashes that are indicative of
memory safety errors in application. This technique has obvious applications
to discovering one class of bug in machine learning systems---crashes---but
can we use fuzzing to find
bugs that silently corrupt the system's outputs?
In this section, we use OpenCV as a running example while describing our bug discovery methodology.

Our use case is, in one sense, a natural fit for general purpose fuzzing
because we can have a single program that runs on some input (i.e. an image)
to produce some output (i.e. a text classification of that image). However, we have
to ensure that we separate and identify both bug types of interest: crashes and silent corruption. To do this we introduce a technique we call
\emph{steered fuzzing}.

We use American Fuzzy Lop (AFL)~\cite{afl} to instrument and fuzz-test machine learning programs.
AFL was designed and is commonly used for finding crashes due to
parsing failures, so the AFL loop involves running an application on multiple inputs and creating a report if an input causes a crash.
AFL utilizes a genetic algorithm to generate inputs while maximizing the code coverage and
has heuristics
to discriminate between unique
crashes and duplicates.
We want to capitalize on AFL's ability to maximize code coverage while also finding crashing inputs.

A \emph{steered fuzzing} workflow begins with a test case with a known
outcome; for example, when analyzing OpenCV, we start with an image that contains a human face. The three outputs
from the program under test might be: \emph{crash}, \emph{negative prediction}, (e.g. no face found) or \emph{positive prediction} (e.g. face found).
The default behavior with the initial test case is to find a face.
Our fuzzing should mutate the image to change the output of the
program under test to \emph{negative prediction} while avoiding \emph{crash}.

When we are searching for
such logical failures, we do not care about inputs that produce crashes when
OpenCV attempts to parse the image (although there is a disturbingly large number
of these inputs). The first part of our \emph{steered fuzzing} technique
brackets the parsing regions of the program in a handler for
the \texttt{SIGSEV} signal. The handler simply \texttt{exit}'s the program
when a segmentation violation occurs. This prevents the crash and
obscures it from AFL, which then believes that the application exited
normally.

We then re-enable crashes in the application and check the outcome of important decision points in the ML algorithm.
For example, we check the result of the face detection step, which corresponds to the outcome of the prediction phase from Figure~\ref{attacksurface}.
If the system failed to find a face, we induce a crash by manually dereferencing an invalid pointer.
In this way, AFL recognizes when it has changed the output of the program to
\emph{no face found} without any change to AFL itself.
Similarly, we can instrument the outputs of each of the components described in Section~\ref{sec:attack-surface}, to check for the presence of exploitable ML bugs.


\section{Results}
\label{sec:results}

We search for exploitable ML bugs in the OpenCV~\cite{opencv} computer vision library and in the Malheur~\cite{malheur} library for analyzing and clustering malware behavior.
We select these libraries because they are open source and they are widely adopted.

OpenCV provides its users with a common framework for computer vision applications and can process still imagery, live streaming video, and previously recorded video clips.
For example, businesses can use computer vision and machine learning to reinforce physical security systems~\cite{homesec}. In such a scenario, an adversary may wish to thwart physical security through attacking the machine learning application itself.

Malheur is a security tool that performs analysis on malware reports that were recorded within sandboxed environments. Malheur can cluster the reports to
determine which samples likely belong to the same malware family;
these malware reports can be raw text files or compressed file archives. Malheur relies on {\tt libarchive} to extract the malware reports from the file archives. An adversary that desires to delay analysis of their malware may target Malheur through crafted file archives and corrupt in-memory data. Data corruption will cause misclustering and allow the adversary to accomplish their goal.

As a result of our research, we responsibly disclosed five vulnerabilities (to which three were assigned  CVE-IDs). The {\tt libarchive} and Malheur system maintainers patched two of the vulnerabilities; as of January 27, 2016, the OpenCV maintainers acknowledged three vulnerabilities and would address the issues in future releases. These vulnerabilities still exist in the current version of OpenCV.
Table~\ref{tsummary} summarizes the vulnerabilities we found and their impact.

\begin{table*}[t]
\begin{tabular}{|p{3.5cm}|p{1.75cm}|p{2.5cm}|p{1.5cm}|p{6.0cm}|}
\hline
{\bf Vulnerability} & {\bf Application} & {\bf CVE-ID} & {\bf Exploited} & {\bf Impact} \\ \hline
Heap Corruption in Feature Extraction & OpenCV & CVE-2016-1516  & \checkmark & Arbitrary code execution via double\_free \\ \hline
Heap Corruption in Feature Extraction & OpenCV & CVE-2016-1517 & \checkmark & Denial of service attack via corrupt\_chunks and segfault \\ \hline
Inconsistent rendering in Feature Extraction & OpenCV & n/a & \checkmark & Partial rendering of {\tt JPG} files results in evasion \\ \hline
Heap Corruption in Feature Extraction & Malheur & CVE-2016-1541 & \checkmark & Arbitrary code execution on all Linux and OS X systems via corrupted archive \\ \hline
Heap Corruption in Feature Extraction & Malheur & GitHub patch & \checkmark & Memory corruption via unsafe bounds checking results in misclustering \\ \hline
Loss of Precision in Feature Extraction & Malheur & n/a & & Loss of precision results in misclustering \\ \hline
Loss of Precision in Model Representation & Scikit-Learn & n/a & \checkmark & Loss of precision results in mispredictions \\
\hline
\end{tabular}
\caption{Summary of ML Hunter findings.}
\label{tsummary}
\end{table*}

\subsection{Discovery Results} \label{fuzzing}
\paragraph{OpenCV}
We discovered bugs in OpenCV's image processing library, and we identified various conditions under which a valid JPG would cause an algorithm to terminate.
Two vulnerabilities (CVE-2016-1516 and CVE-2016-1517) exist in the \emph{feature extraction / selection} portion of the ML attack surface in Figure~\ref{attacksurface} and cause memory corruption when freeing a matrix allocated for image processing. In both CVEs, heap corruptions overflow fields in the matrix object and allow illegal access to memory locations when matrix objects are deallocated. Many examples exist in which an adversary can exploit similar vulnerabilities in image processing code and achieve remote code execution on a victim's system~\cite{projectzero_samsung,stagefright,libpng}. Our third vulnerability exists in OpenCV's custom image rendering library. Its improper handling of file artifacts and partial rendering of particular JPG images prevent consistent image classification.

During the steered fuzzing phase, these vulnerabilities and inconsistencies served as the basis for crafting legitimate input images that evade facial recognition detection. When used as-is, these images induce denial-of-service (DoS) crashes against OpenCV. DoS crashes in such an application require an administrator or operator to manually intervene to bring the system back online. Proof-of-concept SQL injection attacks already exist against video monitoring software that law enforcement organizations use to read license plates and issuing fines~\cite{sqlcar}; one can understand the ramifications of a DoS attack against similar applications or even autonomous driving vehicles using similar computer vision software.

A potentially viable defense against these crash-inducing images starts with first filtering input based on a render-check using the Python Image Library (PIL)~\cite{pythonPIL} and the code snippet in Listing~\ref{lst:PIL}:

\begin{lstlisting}[label={lst:PIL},caption={Image render-check using PIL},language=Python]

from PIL import Image
def is_image_ok(filename):
    try:
        Image.open(filename).load()
        return True
    except:
        return False
\end{lstlisting}
Of the 3197 images we found that induce crashes in OpenCV, PIL only allows 7 images to bypass this filter, resulting in a 0.0022\% false negative rate. Yahoo! Flickr's proprietary image rendering solution allows 6 crash-inducing images through. These crash-inducing images are publicly available for viewing.\footnote{https://www.flickr.com/gp/138669175@N07/L53K8e}

\paragraph{Malheur} We discovered a critical bug within \verb!libarchive! as used by Malheur.
The vulnerability was issued CVE-2016-1541~\cite{malheurcve} and was patched in \verb!libarchive 3.2.0! on May 1, 2016.
This vulnerability affected \emph{every version of Linux and OS X}, given that \verb!libarchive! is pre-packaged in these operating systems for handling various file archives. 
A successful exploit would allow an attacker to achieve arbitrary code execution by exploiting the inconsistent handling of {\tt .tar.gz} compressed archives. 
This vulnerability occurs within the \emph{feature extraction} block within Figure~\ref{attacksurface}, given that the function inherently relies upon \verb!libarchive! for attaining data. Once an attacker achieves arbitrary code execution, they have unlimited influence over the classification of the ML application. 
This bug could trigger another bug in Malheur's feature matrix extraction/selection and was patched on March 6, 2016~\cite{malpatch}. 
A third vulnerability in Malheur results in loss of precision when casting a \emph{double} value to \emph{float}.
Section~\ref{sec:steeredfuzzing} explores the impact of corrupting the feature matrix in greater depth.

\paragraph{Scikit-Learn and NumPy} We also discovered a loss of precision vulnerability in the popular SkLearn Python framework and the underlying NumPy library that, when exploited, results in mispredictions. 
Due to the fact that AFL is not compatible with Python, we manually inspected the source code during bug discovery. 

\subsection{Steered Fuzzing Results}\label{sec:steeredfuzzing}
\begin{figure}[!ht]
\centering
\includegraphics[width=.4\textwidth]{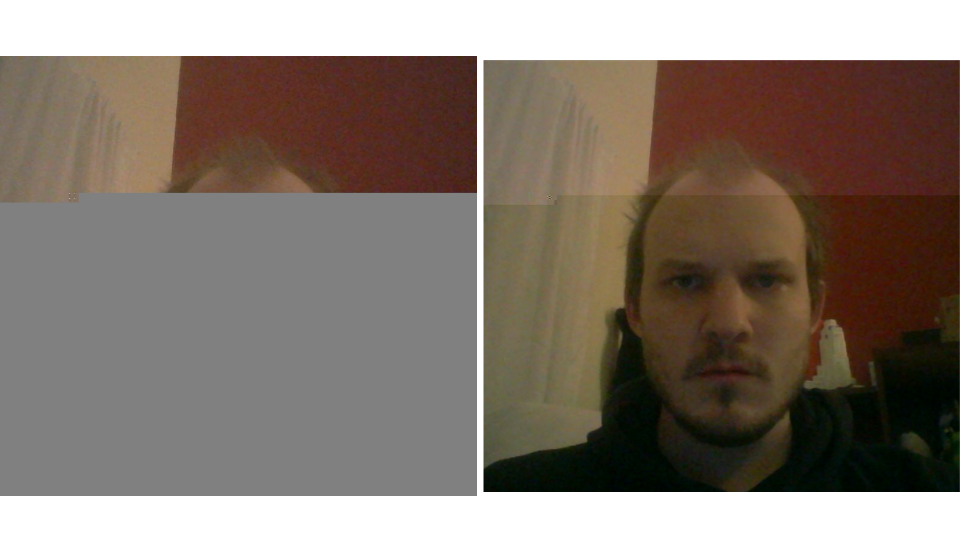}
\caption{OpenCV incorrectly rendering a picture.}
\label{res:misrender}
\end{figure}

\paragraph{OpenCV} An attacker can exploit OpenCV's inconsistent rendering of images to induce silent failures and thwart facial detection within the prediction block of the ML attack surface in Figure~\ref{attacksurface}. To begin, AFL utilizes a seed image with a shoulder-up picture of a person. The source code snippet that performs facial recognition\footnote{\seqsplit{https://github.com/opencv/opencv/blob/a8e5d1d9fdae183762d4e06a7e25473d10ef1974/samples/cpp/facedetect.cpp\#L202}} is a prime candidate for injecting the the logic branch (Listing~\ref{lst:facecheck}) which allows us to induce a crash when the picture of the face is no longer detected.
\begin{lstlisting}[label={lst:facecheck},caption={Logic branch injection for facial detection},language=C]

if(faces.size() == 0) {
    *((int *) 0xdeadbea7) =
    0xdeadbeef;
}
\end{lstlisting}
After 10.1 million permutations of the seed image, AFL crafted Figure \ref{res:misrender}. This image is incorrectly rendered by OpenCV, as seen on the left, but is clearly renderable by Google Photos, as seen on the right.
In five out of five trials, this method successfully recreated photos that exercise this rendering bug.
As these images are correctly formatted JPEG files, they bypass the PIL render-check described in Listing~\ref{lst:PIL}.
In contrast to existing techniques for crafting adversarial samples that evade detection~\cite{szegedy2013intriguing,biggio2013data,biggio2011support,miller2014adversarial,barreno2010security},
our \emph{attack does not depend on the learned model and succeeds from the first attempt}.
This represents a new attack vector against machine learning, illustrating how bugs in ML code can provide a substantial advantage to the attacker.

\paragraph{Malheur} Building on Malheur's inability to handle corrupted archive files, discussed previously, the steered fuzzing technique can corrupt Malheur's feature matrix and induce silent failures in prediction results. Thus, this vulnerability impacts all aspects of \emph{clustering} within the generalized attack surface in Figure~\ref{attacksurface}; an attacker can corrupt the in-memory data for unlabeled samples, tamper with the in-memory feature matrix, and affect the clustering results based on degree of induced skew. The vulnerable line of code\footnote{\seqsplit{https://github.com/rieck/malheur/blob/75ffd2498e964aa7d09782bf5a0d31afde36585f/src/fvec.c\#L382!}} uses the variable \verb!j! which is dependent on user-provided input. Thus, steered fuzzing can craft a corrupted archive file to traverse the heap and stomp over existing values in the feature matrix as shown in Listing~\ref{lst:heapcorrupt}.
\begin{lstlisting}[label={lst:heapcorrupt},caption={Example of directed heap corruption},language=C]

if (((unsigned long)&t[j -1] >
(unsigned long)&fv->val[0]) &&
((unsigned long)&t[j-1] <
(unsigned long)
((unsigned long)&fv->val[0]+
(unsigned long)fv->mem))){
    *((int *) 0xdeadbea7) =
    0xdeadbeef;
}
\end{lstlisting}

As this is a heap corruption vulnerability, we performed our proof-of-concept (PoC) exploit with address space layout randomization turned off. An attacker can couple our PoC exploit with ASLR bypass~\cite{durden2002bypassing} techniques using another information disclosure exploit to find the desired offset. Additionally, our exploit uses a file archive; the exploitation success varies among operating systems and architectures as expected.

Expanding upon this example, an adversary can inject additional logic branches to control the degree in which the corrupted file impacts the feature matrix.
Given enough time, AFL can generate inputs that increasingly skew the clustering of benign files the adversary did not craft.

This represents a second attack vector that provides new capabilities for adversaries.
Unlike prior attacks proposed in the adversarial machine learning literature, this attack introduces the ability to \emph{manipulate the in-memory representation of inputs not provided by the adversary}.
From an adversarial perspective, this attack
provides the opportunity to miscluster benign samples, or other malicious samples, to obfuscate the attacker's own malicious sample.
An adversary can achieve this by inducing false negatives (more stealthy and desired) or false positives (junk reports).
Again, this attack requires only one malicious sample and succeeds from the first attempt, owing to the bug.
The \verb!libarchive 3.2.0! and Malheur GitHub patch rendered this bug unexploitable, as corrupted archives are rejected on ingest.

We also discovered a bug in Malheur that we could not exploit.
During feature normalization in Malheur, both functions\footnote{\seqsplit{https://github.com/rieck/malheur/blob/75ffd2498e964aa7d09782bf5a0d31afde36585f/src/fmath.c\#L37}} \verb!fvec_norm1()! and \verb!fvec_norm2()! return a value of type \verb!double! but it is then normalized to a \verb!float!. Using steered fuzzing, an attacker can discover instances where type casting from \verb!double! to \verb!float! yields a discrepancy.
\begin{lstlisting}[label={lst:precisionloss},caption={Example of discovering precision loss},language=C]

if (abs((double) (f->val[i] / s) -
(float) (f->val[i] / s)) > epsilon){
    *((int *) 0xdeadbea7) = 0xdeadbeef;
}
\end{lstlisting}
The {\tt epsilon} in Listing~\ref{lst:precisionloss} represents the loss of precision an adversary wishes to induce. In this particular instance, steered fuzzing did not discover any value of {\tt epsilon} that caused misclustering. While our attack was unsuccessful, this approach could be a viable attack vector elsewhere.
\par
We discovered that such a vulnerability is present within the Scikit-Learn machine learning library for Python and its underlying reliance on NumPy.
When defining {\tt ndarray} objects from Python lists without explicitly specifying a data type, the library infers the resulting data type according to undocumented heuristics.
The NumPy arrays are used by Scikit-learn during both training and testing of the Linear Regression algorithm.
This attack forces NumPy to set the {\tt ndarray} data type as {\tt object}, which preserves the underlying data type of each element.
The Scikit-learn sanity checks ensure that the training and testing data types match.
Because both the training and the testing arrays are of type \textit{object}, the arrays pass the Scikit-learn checks.
Our proof-of-concept (PoC) code places Python {\tt float} and {\tt long} values in the arrays before that data is ingested by the Scikit-learn module.
When the input numbers are very large, this results in a loss of precision from casting.
Specifically, the PoC shows how the regression model coefficients are drastically changed when using {\tt float} instead of {\tt long} in the training dataset; this results in \emph{mispredictions} at testing.
In absence of a fuzzing tool for Python, we discovered this bug by manually inspecting the Scikit source code, steered by the attack surface guidelines.
These bugs illustrate a third attack vector that potentially enables new adversarial capabilities.

\paragraph{Disclosure experience.}
Three of the vulnerabilities we discovered were assigned new CVE-IDs
(CVE-2016-1516, CVE-2016-1517,and CVE-2016-1541),
as they enabled arbitrary code execution or denial of service attacks.
Disclosing these vulnerabilities allowed us to draw an interesting comparison between the community's reaction to these bugs and the ML-specific bugs we introduce in this paper.
In particular, the bugs that led to misprediction, misclustering, or model divergence---including a Malheur memory corruption bug that allows the adversary to control the feature matrix, but not to inject arbitrary code---did not receive CVE numbers.
Many of these bugs were labeled WONTFIX.
This emphasizes the fact that ML bugs are currently a misunderstood threat.


\section{Related work}
\label{sec:related}

This section presents prior work on fuzzing and adversarial machine learning. Adversarial machine learning research focuses on crafting adversarial samples. The key distinction in our work is that we exploit bugs in machine learning code that give the adversary an advantage in conducting these attacks. 

Insufficient input sanitization is a common cause of exploitable bugs~\cite{mcgraw2006software}.
Fuzzing is an automated technique that allows developers to test how gracefully their application handle various valid and invalid input~\cite{DBLP:journals/cacm/MillerFS90, oehlert2005violating}. Fuzzers assist developers with isolating potentially buggy code and can play a critical role in identifying locations in need of input sanitization.

The field of adversarial machine learning has developed several methods for attacking ML systems, typically by querying ML models.
Barreno et al. \cite{barreno2010security} proposed a general classification system for these attacks.
Integrity attacks allow hostile input into a system and availability attacks prevent benign input from entering a system.
Concept drift \cite{tsymbal2004problem} is a phenomenon that occurs within machine learning systems as the prediction becomes less accurate over time due to unforeseen changes. Identifying concept drift, whether sudden or gradual, can be difficult in the presence of noise. Ideally, machine learning systems should combine robustness to noise and sensitivity to concept drift. Adversarial drift \cite{miller2014adversarial} describes intentionally induced concept drift in an effort to decrease the classification accuracy. Biggio et al. \cite{biggio2013data} described a threat model in which an attacker desires to conceal malicious input in an effort to evade detection without negatively impacting the classification of legitimate samples. According to Biggio, an attacker may wish to inject malicious input to subvert the clustering process, rendering the resulting knowledge useless. The adversarial classifier reverse engineering~\cite{lowd2005adversarial} describes techniques for learning sufficient information about a classifier to instrument adversarial attacks. This information provides attackers and defenders with an understanding of how susceptible their system is to external adversarial influence. Newsome et al.~\cite{newsome2006paragraph} introduce a delusive adversary that provides malicious input in an attempt to obstruct the ML training phase; the attacker assumes full control over the input and its order.

In an analysis of a neural network trained for image processing tasks, Szegedy et al. \cite{szegedy2013intriguing} identified that an adversary can apply a perturbation to an image that is imperceptible to humans yet it changes the network's prediction. 
Goodfellow et al. \cite{goodfellow2014explaining} present a fast method for generating adversarial perturbations to fool an image classifier.
Research from Cha et al. \cite{cha2015program} further explores automated generation of such perturbations. Utilizing a well-formed seed input, a mutational fuzzer iteratively manipulates the seed to achieve maximum path traversal in a target program. This technique can isolate particular sets of input that cause the program to enter a state that might be of interest for an attacker.

Cretu et al.~\cite{cretu2008casting} discuss the process and importance of ``casting out demons,'' sanitizing ML training datasets for anomaly detection (AD) sensors. AD systems inherently receive malicious input and anomalous events that may drastically impact the system's tuning and instrumentation. Accounting for data that may negatively impact the accuracy of the system's classifier can enhance its overall robustness.


\section{Discussion and Future Work}
\label{sec:discussion}

In this paper, we focus on attacks against machine learning systems.
However, our threat model has a broader applicability.
For example, nation-state adversaries might be interested in attacking long-running simulations on supercomputers, with the aim of subtly skewing their results.
In high performance computing, outputs are generally difficult to validate and expensive to re-compute, so it is difficult to defend against such attacks.
Other data analytics systems may also be susceptible to such attacks.

For some of the bugs that we discovered, it is unclear who is responsible for fixing them.
Should the Malheur maintainers have to worry about bugs in \verb!libarchive! in order to preserve the integrity of their application? Should the architects of OpenCV sacrifice performance for the sake of handling invalid input that developers did not filter? As more and more everyday devices begin to incorporate ML processing, this ambiguity must be explicitly resolved in order to provide secure systems.

Section~\ref{sec:methods} describes a semi-automated approach for discovering bugs in machine learning platforms through categorizing the backtrace of crash-inducing results. Tools such as {\tt !exploitable}~\cite{exploitable} provide researchers with automated crash analysis and the likelihood that the crash is exploitable. Overlaying the findings from such a tool on top of our generalized attack surface could expedite the discovery phase.

Section~\ref{sec:results} explores many techniques that, at first glance, are only feasible because the targeted source code is publicly available. Recently, Papernot et al.~\cite{DBLP:journals/corr/PapernotMGJCS16} proposed model extraction attacks, by building surrogate classifiers that approximate black-box ML models. A logical next step in expanding our research would be understanding the overlap between building substitution models of proprietary classifiers and unique edge cases that result in bugs in the black box system.

An adversary discovering the possibility of ``linchpin values'' that appear during feature matrix construction would be another decisive shift towards an attacker's influence on ML systems. Linchpin values are consistent ranges of values within a feature matrix, that when present, result in a specific classification. Ribeiro et al.~\cite{ribeiro2016should} proposed a technique for model explanation by building locally optimal classifiers around points of interest. Building upon their work, researchers may apply various analytic techniques to determine if there are common values or thresholds within a feature matrix that, when present, always result in a certain classification. With this information, an attacker could use steered fuzzing to craft arbitrary input that would guarantee a misclassification in the targeted system.

\section{Conclusions}
\label{sec:conclusions}
Entities that choose to trust data from unvetted sources subject themselves to a plethora of potential attacks in which a miscreant only requires minimal control over the entire dataset. For an attacker that wishes to control the decision-making process of its competitors or adversaries, this represents a powerful paradigm shift in attack vectors. We discovered several vulnerabilities within OpenCV and Malheur that allow an attacker to exploit bugs in underlying dependencies and the applications themselves to gain a marked advantage in influencing or out-right controlling the output of ML applications.

\bibliographystyle{abbrv}
\bibliography{security.bib}

\end{document}